\documentclass[preprint,superscriptaddress]{revtex4-1}

\usepackage{graphicx}
\usepackage[ansinew]{inputenc}
\usepackage{array}
\usepackage{color}
\usepackage{amsmath}
\usepackage{amsxtra}
\usepackage{amstext}
\usepackage{amssymb}

\usepackage{latexsym}
\usepackage{dsfont}

\begin{document}
\renewcommand{\figurename}{Figure}

\title{Emergence of Superlattice Dirac Points in Graphene on Hexagonal Boron Nitride}

\author{Matthew Yankowitz}
\author{Jiamin Xue}
\author{Daniel Cormode}
\affiliation{Physics Department, University of Arizona, 1118 E 4th Street, Tucson, AZ 85721, USA}
\author{Javier D. Sanchez-Yamagishi}
\affiliation{Department of Physics, Massachusetts Institute of Technology, Cambridge, MA 02138, USA}
\author{K. Watanabe}
\author{T. Taniguchi}
\affiliation{Advanced Materials Laboratory, National Institute for Materials Science, 1-1 Namiki, Tsukuba 305-0044, Japan}
\author{Pablo Jarillo-Herrero}
\affiliation{Department of Physics, Massachusetts Institute of Technology, Cambridge, MA 02138, USA}
\author{Philippe Jacquod}
\affiliation{Physics Department, University of Arizona, 1118 E 4th Street, Tucson, AZ 85721, USA}
\affiliation{D\'epartment de Physique Th\'eorique, Universit\'e de Gen\`eve CH-1211 Gen\`eve, Switzerland}
\author{Brian J. LeRoy}
\email{leroy@physics.arizona.edu}
\affiliation{Physics Department, University of Arizona, 1118 E 4th Street, Tucson, AZ 85721, USA}

\date{\today}

\begin{abstract}
The Schr\"odinger equation dictates that the propagation of nearly free electrons through a weak periodic potential results in the opening of band gaps near points of the reciprocal lattice known as Brillouin zone boundaries~\cite{Ashcroft}. However, in the case of massless Dirac fermions, it has been predicted that the chirality of the charge carriers prevents the opening of a band gap and instead new Dirac points appear in the electronic structure of the material~\cite{Park2008a, Park2008b}. Graphene on hexagonal boron nitride (hBN) exhibits a rotation dependent Moir\'e pattern~\cite{Xue2011,Decker2011}.  In this letter, we show experimentally and theoretically that this Moir\'e pattern acts as a weak periodic potential and thereby
leads to the emergence of a new set of Dirac points at an energy determined by its wavelength.  The new massless Dirac fermions generated at these superlattice Dirac points are characterized by a significantly reduced Fermi velocity. The local density of states near these Dirac cones exhibits hexagonal modulations indicating an anisotropic Fermi velocity.
\end{abstract}

%\pacs{73.23.-b, 72.25.Dc, 85.75.-d} 

\maketitle

Due to its hexagonal lattice structure with a diatomic unit cell, graphene has 
low-energy electronic
properties that are governed by the massless Dirac equation~\cite{CastroNeto2009}. 
This has a number of consequences, among them Klein tunneling~\cite{Katsnelson2006, Klein1929,Stander2009, Young2009}
which prevents electrostatic confinement of charge carriers and inhibits the fabrication of standard semiconductor devices. 
This has motivated a number of recent theoretical investigations of graphene in periodic potentials~\cite{Park2008a,Park2008b,Barbier2008,Brey2009,Brey2010,Burset2011,Ortix2011}, which
explored ways of controlling the propagation of charge carriers by means of
various superlattice potentials. 
On the analytical side, one-dimensional potentials
render particle propagation anisotropic~\cite{Park2008a, Park2008b,Barbier2008,Burset2011} and 
generate new Dirac points, where the electron and hole bands meet, at energies $\pm \hbar v_{\rm F} |{\bf G}|/2$ given by the
reciprocal superlattice vectors ${\bf G}$~\cite{Park2008a, Park2008b}, and at zero energy
for potentials with strengths exceeding $\approx \hbar v_{\rm F} |{\bf G}|$~\cite{Brey2009}.
Numerical approaches have extended several of these results  
to the case of two-dimensional potentials~\cite{Park2008a,Park2008b,Burset2011,Ortix2011}.
Unlike for Schr\"odinger fermions, the periodic potentials generally induce new Dirac points but
do not open gaps in graphene, owing to the chiral nature of the
Dirac fermions.

Recent STM topography experiments have reported well developed 
Moir\'e patterns in graphene on crystalline substrates, which suggests that the latter
generate effective periodic potentials~\cite{Xue2011,Decker2011,Marchini2007,Parga2008}. Of 
particular interest is hexagonal boron nitride (hBN), because it is an insulator which only couples weakly to graphene.  Furthermore, graphene on hBN exhibits the highest mobility
ever reported for graphene on any substrate~\cite{Dean2010}, and has strongly suppressed
charge inhomogeneities~\cite{Xue2011,Decker2011}. 
Hexagonal boron nitride is a layered material whose planes have the same atomic structure
as graphene, with a 1.8\% longer lattice constant. The influence of the weak graphene-substrate interlayer coupling on the electronic transport and spectroscopic properties of graphene is not well understood. In particular, there is to date no theory for local electronic properties such as those probed in STM experiments. Below we show that
periodic interlayer couplings generate a new Dirac point at an energy determined by the wavevector of the periodic potential.  The presence of this new Dirac point is reflected in two dips in the density of states, symmetrically placed at $E= \pm \hbar v_{\rm F} |{\bf G}|/2$ around the $E=0$ graphene Dirac point but generally of asymmetric strength.  There is also a periodic modulation of the local density of states, indicating an anisotropic Fermi velocity, with the same period as the superlattice topographic Moir\'e pattern.

\begin{figure}[t]
\includegraphics[width=12cm]{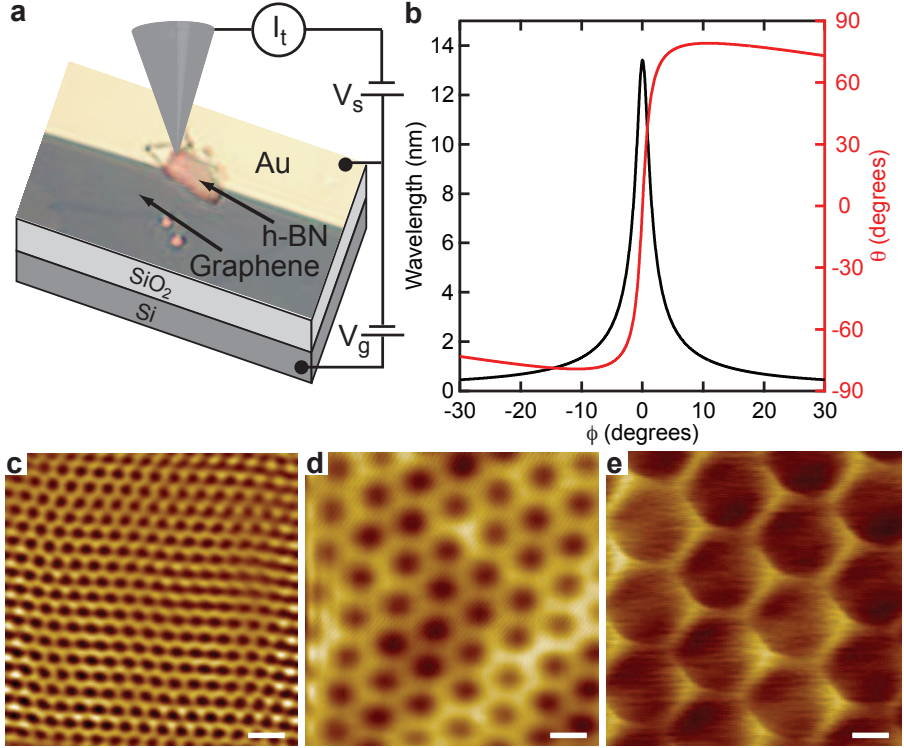} 
\caption{Graphene device schematic and STM Moir\'e images. (a) Schematic of the measurement setup showing the STM tip and an optical microscope image of one of the measured samples.  (b) Superlattice wavelength $\lambda$ (black) and rotation (red) as a function of the angle between the graphene and hBN lattices. (c)-(e) STM topography images showing (c) 2.4 nm, (d) 6.0 nm and (e) 11.5 nm Moir\'e patterns.  Typical imaging parameters were sample voltages between 0.3 V and 0.5 V and tunnel currents between 100 pA and 150 pA.  The scale bars in all images are 5 nm.}
\label{fig:topography}
\end{figure}

The fabrication procedure used for creating the graphene on hBN devices results in a random rotational orientation between the graphene and hBN lattices.  This rotation between the lattices and the longer lattice constant for hBN leads to topographic Moir\'e patterns.  Given the lattice mismatch $\delta$ between hBN and graphene, the relative rotation angle $\phi$ between the two lattices uniquely determines the Moir\'e wavelength $\lambda$ as
\begin{equation}\label{eq:lambda}
\lambda = \frac{(1+\delta)a}{\sqrt{2(1+\delta)(1-\cos(\phi))+\delta^2}} \, ,
\end{equation}
where $a$ is the graphene lattice constant.  The relative rotation angle $\theta$ of the Moir\'e pattern with respect to the graphene lattice is given by
\begin{equation}\label{eq:theta}
\tan(\theta) = \frac{\sin(\phi)}{(1+\delta)-\cos(\phi)} \, .
\end{equation}
Figure ~\ref{fig:topography}(b) plots the wavelength of the Moir\'e pattern (black) and rotation angle (red) as a function of $\phi$.  Due to the lattice mismatch, there is a Moir\'e pattern for all orientations of graphene on hBN with a maximum possible length of about 14 nm.  Figs.~\ref{fig:topography} (c)-(e) show STM topography images of Moir\'e patterns for three different rotations of the graphene lattice.

%%%%%%%%%%%%%%%
%%%%%%%%%%%%%%%%%%

We explore how this Moir\'e structure influences the local density of states (LDOS) $\rho({\bf r},E) = \sum_n |\psi_n({\bf r})|^2 \delta(E-E_n)$ in the graphene layer. This quantity dominates dI/dV STM measurements as long as the density of states of the STM tip and the tunneling rate from the tip to the sample are constant.  We take a lattice Hamiltonian for a graphene monolayer deposited on top of a hBN monolayer, where the two layers are rotated with respect to one another, and the spatially dependent interlayer hopping $t_\perp$ is calculated from nearest neighbor and next nearest neighbor interlayer coupling~\cite{Xue2011}.  We numerically calculate $\rho({\bf r},E)$ for this model using the Lanczos method~\cite{Dagotto1994} (see Supplementary Information). 

\begin{figure}[h]
\newpage
\includegraphics[width=12cm]{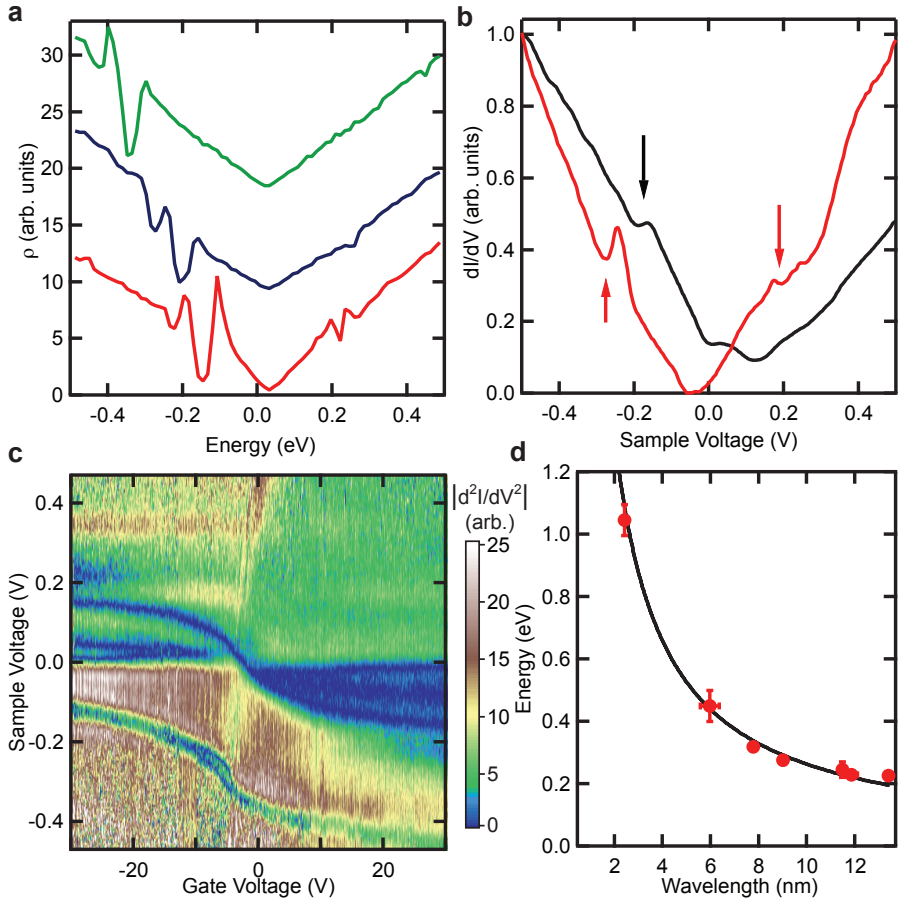} 
\caption{Density of states of graphene on hBN showing new superlattice Dirac point. (a) Theoretical LDOS curves for three different rotation angles between graphene and hBN, red is $\phi=0.5^o$ (12.5 nm), blue is $\phi=1^o$ (10.0 nm) and green is $\phi=2^o$ (6.3 nm).  The curves have been vertically offset for clarity.  (b) Experimental dI/dV curves for two different Moir\'e wavelengths, 9.0 nm (black) and 13.4 nm (red).  The dips in the dI/dV curves are marked by arrows. (c) $|$d$^2$I/dV$^2|$ as a function of gate and sample voltage for the 9.0 nm Moir\'e pattern showing the shift of the Dirac point and one of the dips. (d) Energy of the dips away from the Dirac point as a function of Moir\'e wavelength.  The red points are the experimentally measured values and the black line is the expected theoretical dependence.}
\label{fig:dI-dV}
\end{figure}

Dips in the calculated $\rho({\bf r},E)$ are clearly seen in Fig.~\ref{fig:dI-dV}(a).  The energy of these dips changes as a function of the rotation angle $\phi$ and hence the Moir\'e wavelength.  We have also observed the dips in the experimental dI/dV curves as shown in Fig.~\ref{fig:dI-dV}(b).  The black curve is for a 9.0 nm Moir\'e pattern and the energy of the dip is 0.28 eV from the Dirac point.  The red curve is for a 13.4 nm Moir\'e pattern and the energy of the dips decreases to 0.22 eV from the Dirac point.  Both experimentally and theoretically, we found that the relative strength of the dips in the conduction and valence band are different, with the dip in the valence band being much deeper than the dip in the conduction band.  In our numerical calculations, we identified that most of this asymmetry arises due to next nearest neighbor interlayer coupling, which effectively induces modulated hopping between different graphene sublattices and breaks electron-hole symmetry (see Supplementary Information). Figure ~\ref{fig:dI-dV}(c) plots $|$d$^2$I/dV$^2|$ for the 9.0 nm Moir\'e pattern as a function of gate voltage and sample voltage.  We clearly see the Dirac point in this measurement crossing the Fermi energy near zero gate voltage.  There is a second dip which moves parallel to it that is offset by -0.28 V in sample voltage.  This dip is due to the superlattice periodic potential induced by the hBN and indicates the emergence of a new superlattice Dirac point.

We have observed these dips in the LDOS for seven different Moir\'e wavelengths.  The energy of the dips from the Dirac point is plotted (red points) as a function of wavelength in Fig.~\ref{fig:dI-dV}(d).  The solid black line plots the expected energy dependence $E = \hbar v_{\rm F}|{\bf G}|/2 = 2\pi \hbar v_{\rm F} /\sqrt{3}\lambda$ assuming the linear band structure of graphene and $v_F=1.1 \times 10^6$ m/s.  For the necessary high resolution spectroscopy, our STM is limited to observing dips in an energy range of $\sim\pm$1 V which restricts the Moir\'e wavelengths to longer than 2 nm.

%%%%%%%%%%%%%%%
%%%%%%%%%%%%%%%%%%

To better understand these dips, we focus on the low-energy regime and neglect intervalley scattering in graphene. This is justified by the energy range in our STM experiments and the long wavelength of the Moir\'e potential. The interlayer hopping term between the graphene and hBN layers reflects the same periodic structure as the Moir\'e pattern~\cite{Xue2011}. Therefore we model the influence of the hBN by an effective periodic potential with the same symmetry as the observed Moir\'e pattern. 
We accordingly consider the single-valley Hamiltonian
\begin{equation}\label{Hgraphene}
\hat{H} = \hbar v_{\rm F} {\bf k} \cdot \vec{\sigma}  + V \sum_\alpha \cos({\bf G}_\alpha
{\bf x})  \, {\cal I} \, ,
\end{equation}
where ${\bf k}=(k_x,k_y)$, $\vec{\sigma}$ is a vector of Pauli matrices and ${\cal I}$ is the identity matrix. The potential strength is estimated as $V = 0.06$ eV from numerical second-order perturbation theory, and the ${\bf G}_\alpha$ are the reciprocal superlattice vectors corresponding to the periodic potential generated by the hBN substrate. The reciprocal superlattice vector ${\bf G}_1 = (4 \pi/\sqrt{3} \lambda) (\cos \theta, \sin \theta)$ is determined by the relative rotation of the graphene and hBN lattices according to Eqs.~(\ref{eq:lambda}) and (\ref{eq:theta}).  The other superlattice wavevectors are obtained by two rotations of 60$^{\circ}$.  Larger superlattice
vectors are not included in our model, since the corresponding couplings 
are smaller by more than one order
of magnitude~\cite{Xue2011}.

The dips in $\rho({\bf r},E)$ are due to ${\bf k} \rightarrow -{\bf k}$ processes induced by the periodic potential for values $2 {\bf k} = {\bf G}_\alpha$ corresponding to one of the reciprocal  superlattice vectors ${\bf G}_\alpha$. Unlike for Schr\"odinger particles, the chirality of the Dirac fermions prevents such processes from opening a gap at the edges of the superlattice Brillouin zone, as long as the potential does not break sublattice symmetry (see Supplementary Information).

\begin{figure}[h]
\newpage
\includegraphics[width=12cm]{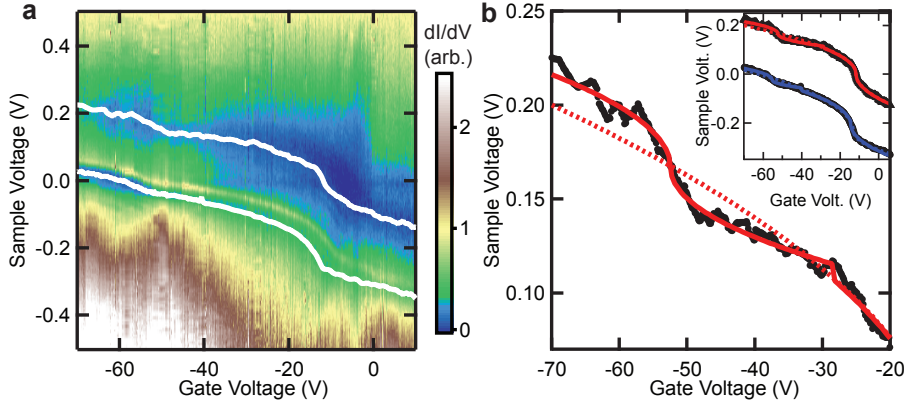} 
\caption{Gate dependence of graphene density of states near superlattice Dirac point. (a) dI/dV as a function of sample and gate voltage showing both the Dirac point and the new superlattice Dirac point.  The white lines mark the location of the Dirac point and superlattice Dirac point.  (b) Shift of the Dirac point as a function of gate voltage when the superlattice Dirac point crosses the Fermi energy (black line).  The solid red line is a theoretical fit for the shift of the Dirac point with the presence of the superlattice Dirac point.  The dashed red line shows the expected shift without the superlattice Dirac point.  The inset shows the shift of both Dirac points over a large gate voltage range as well as theoretical fits (red and blue solid lines).}
\label{fig:NewDP}
\end{figure}

The presence of the new superlattice Dirac points can be detected by examining the gate dependence of the LDOS.  As seen in Fig.  ~\ref{fig:dI-dV}(c), the two Dirac points move in parallel with gate voltage.  Figure  ~\ref{fig:NewDP}(a) plots dI/dV for a 13.4 nm Moir\'e pattern over a larger range of gate voltage.  The white lines show the energy of the two Dirac points as a function of gate voltage.  When the superlattice Dirac point approaches the Fermi energy, the gate dependence of the Dirac points changes.  This is plotted in Fig. ~\ref{fig:NewDP}(b), which tracks the energy of the main Dirac point when the superlattice Dirac point crosses the Fermi energy at -53 V.  Both Dirac points move more quickly with gate voltage, indicating a reduced density of states at this energy.  The gate dependence for the main Dirac point can be fit with the equation $E_{\rm D} =\hbar v_{\rm F}^* \sqrt{\alpha\pi(V_g - V_o)/3}$ where $V_g$ is the gate voltage, $V_0$ is the offset voltage and $\alpha$ is the coupling to the gate.  The factor of three comes from the fact that the periodic potential creates three superlattice Dirac points corresponding to the three reciprocal lattice vectors ${\bf G}_\alpha$.  From the fit, we extract the value of the Fermi velocity at the Dirac point to be $0.94\pm0.02 \times 10^6$ m/s for both electrons and holes.  At the new superlattice Dirac points, the Fermi velocity is reduced to $0.64\pm0.03 \times 10^6$ m/s for the new electrons and $0.78\pm0.03 \times 10^6$ m/s for the new holes.  The reduction agrees with recent theoretical predictions~\cite{Ortix2011} and our numerical calculations.  Evidence for the presence of the superlattice Dirac point can also be seen in the global conductivity as a function of gate voltage (see Supplementary Information).

%%%%%%%%%%%%%%%
%%%%%%%%%%%%%%%%%%

\begin{figure}[h]
\newpage
\includegraphics[width=12cm]{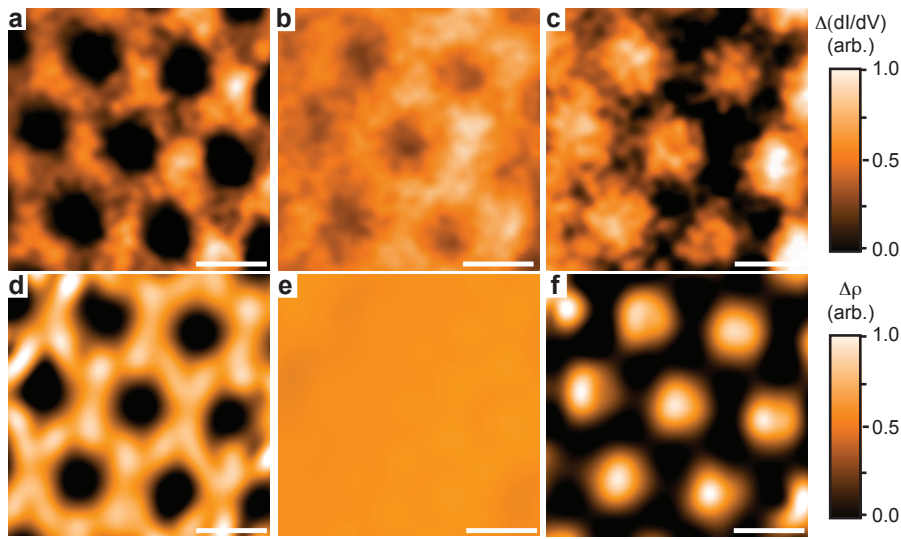} 
\caption{Experimental and theoretical images of LDOS for long wavelength Moir\'e pattern. (a)-(c) Experimental dI/dV maps for a 13.4 nm Moir\'e pattern.  The sample voltages are (a) -0.16 V, (b) 0.17 V and (c) 0.44 V.  The sample voltage in (b) is near the Dirac point since the gate voltage was 60 V while the other two maps are near the energy of the superlattice Dirac points.  (d)-(f) Theoretical dI/dV maps for a 13.4 nm Moire pattern.  The energies are (d) -0.3 eV, (e) 0.03 eV and (f) 0.3 eV.  The energy in (e) corresponds to the Dirac point.  The scale bars in all images are 10 nm.}
\label{fig:moire_scan}
\end{figure}

In addition to the formation of superlattice Dirac points, the presence of the periodic potential also leads to a spatial variation in the LDOS.  Figure ~\ref{fig:moire_scan} shows experimental and numerical images of the LDOS as a function of energy.  Near the Dirac point, (Figs.~\ref{fig:moire_scan} (b) and (e)), we observe a nearly featureless density of states.  This is in agreement with previous STM measurements which show a strong suppression of charge fluctuations in graphene on hBN~\cite{Xue2011,Decker2011}.  At higher energies, the presence of the Moir\'e potential manifests itself as a local variation in the density of states.  Figures ~\ref{fig:moire_scan} (a) and (d) are taken at a lower energy than the superlattice Dirac point in the valence band.  At this energy the hexagonal pattern of the potential is clearly visible.  On the conduction band side, Figs. ~\ref{fig:moire_scan} (c) and (f), the Moir\'e pattern is once again visible but its contrast is inverted.  The centers of the hexagons now correspond to points of increased density of states.  These LDOS modulations indicate that the Fermi velocity is spatially dependent.  Moreover, the inversion indicates that the relative strength of the potential locally determines whether electrons or holes have a higher Fermi velocity.   We observe that the Moir\'e pattern becomes much more visible in the local density of states at energies above the superlattice Dirac points.  Previous measurements of the LDOS in graphene on hBN did not observe any variations due to the short Moir\'e patterns and low energies probed~\cite{Xue2011}.

%%%%%%%%%%%%%%%
%%%%%%%%%%%%%%%%%%

Graphene on hBN devices are becoming widely used due to their improved mobility and reduced charged impurities.  We have shown that lattice mismatch and relative rotation between the graphene and hBN leads to a periodic potential for graphene charge carriers.  This potential creates a new Dirac point whose energy is determined by the wavelength of the potential.  This superlattice Dirac point has the potential to control the transport properties of electrons in graphene as it induces spatially modulated velocities for the charge carriers.  Future work is necessary to exploit this periodic potential for the creation of novel graphene devices.

\section*{Methods}
Graphene on hBN devices were fabricated using two different methods.  In the first method, mechanically exfoliated graphene was transferred to high quality single crystals of hBN which were mechanically exfoliated on a SiO$_2$ substrate~\cite{Dean2010, Xue2011}.  In the second method, commercially available hBN (Momentive AC6004) was exfoliated on SiO$_2$ substrates and then CVD grown graphene was deposited over the hBN.  Both types of devices gave similar results so we do not distinguish between the two types.  After depositing the graphene on hBN, Cr/Au electrodes were written using electron beam lithography.  The devices were annealed at 350$^{\circ}$C for 2 hours in a mixture of Argon and Hydrogen and then at 300 $^{\circ}$C for 1 hour in air before being transferred to the UHV LT-STM for topographic and spectroscopic measurements.  We have measured a total of 29 samples of which 7 had long enough Moir\'e patterns for the observation of a superlattice Dirac point.  

Figure 1a shows a schematic diagram of the measurement setup used for imaging and spectroscopy of the graphene flakes.  All the measurements were performed in UHV at a temperature of 4.5 K.  dI/dV measurements were acquired by turning off the feedback circuit and adding a small (5-10 mV) ac voltage at 563 Hz to the sample voltage.  The current was measured by lock-in detection.

\section*{Acknowledgements}
The work at Arizona was partially supported by the U. S. Army Research Laboratory and the U. S. Army Research Office under contract/grant number W911NF-09-1-0333 and the National Science Foundation CAREER award DMR-0953784, EECS-0925152 and DMR-0706319.  J.S-Y. and P.J-H. were primarily supported by the US Department of Energy, Office of Basic Energy Sciences, Division of Materials Sciences and Engineering under Award DE-SC0001819 and partly by the 2009 US Office of Naval Research Multi University Research Initiative (MURI) on Graphene Advanced Terahertz Engineering (Gate) at MIT, Harvard and Boston University.  P.J. acknowledges the support of the Swiss Center of Excellence MANEP.

\section*{Author contributions}
M.Y., J.X., D.C. and B.J.L. performed the STM experiments of the graphene on hBN.  M.Y. and D.C. fabricated the CVD graphene devices.  J. S.-Y. fabricated the devices on single crystal hBN.  K.W. and T.T. provided the single crystal hBN.  P.J. performed the theoretical calculations.  P. J.-H. and B.J.L. conceived and provided advice on the experiments.  All authors participated in the data discussion and writing of the manuscript.

\section*{Competing financial interests}
The authors declare no competing financial interests.
\newpage

\setcounter{figure}{0}
\renewcommand{\thefigure}{S\arabic{figure}}

\section*{Supplementary Information}

\section{Low energy perturbation theory}

Scattering at the Brillouin zone boundaries due to the periodic potential does not open 
a gap for chiral massless Dirac fermions
as long as sublattice symmetry is preserved.  This can be seen by performing a unitary transformation $\hat{H} \rightarrow \hat{H}' = U_1^\dagger \hat{H} U_1$ on the Hamiltonian of Eq.~(3), with $U_1 = \exp[i \vec{\Lambda}({\bf r}) \cdot \vec{\sigma}]$. When $V/\hbar v_{\rm F} |{\bf G}| < 1$ the linear part of the potential can be gauged out of the Hamiltonian for $\hbar v_{\rm F} \nabla \cdot \vec{\Lambda} = -\hat{V}$. For $V/\hbar v_{\rm F} |G| > 1$, the gauge transformation can no longer remove the linear part of the potential, and the potential creates new Dirac points but still does not open a gap.~\cite{Park2008c,Park2008d,Brey2009a,Brey2010a}.

The magnitude of the dips in the local density of states (LDOS)
 is determined by the strength and the wavelength of the Moir\'e periodic potential. We see this perturbatively by projecting the Hamiltonian of %Eq.~(\ref{Hgraphene}) 
Eq.~(3) onto pairs of eigenstates
$$\psi_{s,{\bf k}} = (1,s \exp[i \theta_{\bf k}])^\dagger \exp[i {\bf k} {\bf r}]/\sqrt{2 \Omega}$$
of $\hat{H}_0 = \hbar v_{\rm F} {\bf k} \cdot \vec{\sigma}$, with the lattice area $\Omega$ and the angle $\theta_{\bf k} = \arctan(k_y/k_x)$. The resulting $2 \times 2$ Hamiltonian reads
\begin{equation}
\hat{H}_{\rm red} = \left(
\begin{array}{cc}
s \hbar v_{\rm F} k/2 & V_{k,k'} \\
(V_{k,k'})^* & s \hbar v_{\rm F} k'/2
\end{array}
\right) \, ,
\end{equation}
with $V_{k,k'} = (V/4) (1+\exp[i(\theta_{{\bf k}'}-\theta_{\bf k})]) \sum_\alpha
(\delta_{{\bf k},{\bf G}_\alpha+{\bf k}'} +\delta_{{\bf k},-{\bf G}_\alpha+{\bf k}'})$. We focus on momenta ${\bf k}={\bf G}_\alpha/2+\delta {\bf k}$, ${\bf k}'=-{\bf G}_\alpha/2+\delta {\bf k}$. For this choice of pairs of eigenstates, it is straightforward to see that  the eigenvalues of $\hat{H}_{\rm red}$ are $E_\pm = (\varepsilon_+ + \varepsilon_-)/2 \pm \sqrt{(\varepsilon_+ - \varepsilon_-)^2/2+ 4 V^2 \delta k_\perp^2/G^2}/4 +{\cal O}(\delta k^3/G^3)$, with $\varepsilon_\pm = \hbar v_{\rm F}  |{\bf G}_\alpha/2 \pm \delta {\bf k}|$, and the component $\delta k_\perp$ of $\delta {\bf k}$ that is perpendicular to ${\bf G}_\alpha$. In particular, there is no energy shift for $\delta {\bf k} \parallel {\bf G}_\alpha$ and a maximal shift for $\delta {\bf k} \cdot {\bf G}=0$. This perturbative result is consistent with the numerical and experimental facts reported here that (i) no gap is opened, neither in the numerically obtained $\rho(E,{\bf r})$, nor in the experimentally obtained STM $dI/dV$, (ii) a 
reduction of (dip in) the density of states is observed at an energy corresponding to $\hbar v_{\rm F} G/2$, and (iii) both the spectroscopic Moir\'e pattern and the dips tend to disappear for periodic potentials with shorter wavelength (hence larger G), when the relative rotation between the hBN substrate and the graphene layer is larger. 
The reduction of the density of states occurs around three points determined by the three 
reciprocal lattice vectors. 
At this level, there is no
$s$-dependence of the strength of the dip, the latter is the same for $s=1$ ($E>0$) as
for $s=-1$ ($E<0$). An $s$-dependence emerges 
once off-diagonal terms $\sim V' \sigma^{x,y}$ 
are included in the Hamiltonian of Eq.~(3).%Eq.~(\ref{Hgraphene}) 

\section{Bilayer lattice Hamiltonian for graphene on hexagonal Boron Nitride}

We consider a lattice Hamiltonian for a graphene monolayer on a single layer of hBN.
For each layer, 
the Hamiltonian reads
\begin{eqnarray}
{\cal H}_\alpha = \sum_i \left[ \epsilon_A(\alpha)  a^\dagger_i(\alpha) a_i(\alpha)
 + \epsilon_B(\alpha)  b^\dagger_i(\alpha) b_i (\alpha) \right]
-t_\alpha \sum_{\langle i , j \rangle} (a^\dagger_i(\alpha) b_j(\alpha) + h.c.) \, ,
\end{eqnarray}
where $a^\dagger_i(\alpha)$ [$a(\alpha)$] and $b^\dagger_i(\alpha)$ [$b(\alpha)$]
are creation [destruction] operators on sublattice A and B, respectively, 
of the graphene ($\alpha=1$)
or hBN ($\alpha=2$) honeycomb lattice, and $\langle i , j \rangle$ indicates that the
sum runs only over nearest neighbors. On the hBN lattice, we choose the
Boron atoms to be on the A
sublattice and the Nitrogen atoms to be on the B sublattice. The on-site energies
are $\epsilon_A(1)=\epsilon_B(1)=0$, $\epsilon_A(2)=3.34$ eV
and $\epsilon_B(2)=-1.4$ eV, and 
the hopping integrals are
$t_1=3.16$ eV and $t_2=2.79$ eV~\cite{Xue2011a} (the precise value of the latter is
of little importance).

We restrict the interlayer hopping potential to nearest-neighbor and next-nearest-neighbor hopping. The interlayer hopping is given by
\begin{eqnarray}
t_{ij}' (m,n) & = & \gamma_\perp \exp[-|{\bf r}_i(m) -{\bf r}_j(n)|/\xi] \, f_{ij} \, ,
\end{eqnarray}
with a characteristic function $f_{ij}=1$ if site $i$ of sublattice $m$ on the graphene (hBN)
sheet 
is nearest or next-nearest neighbor to site
$j$ of sublattice $n$ on the hBN (graphene) sheet, and $f_{ij}=0$ otherwise.  
The parameters $\gamma_\perp=0.39$ eV and $\xi=0.032$ nm are calibrated to fit the interlayer couplings in bilayer graphene\cite{Zhang2008}. 
The two lattices are rotated with respect
to one another by the angle $\phi$, and for each site we determine its nearest and next
nearest neighbor site on the other sheet, numerically
evaluate the distance between the sites and finally the corresponding interlayer hopping.
While $\gamma_\perp$ should in 
principle depend on the sublattice index in the hBN layer, we neglect this dependence
here.
The interlayer coupling Hamiltonian is then given by
\begin{eqnarray}
{\cal H}_\perp &=& -\sum_{ij} \left[ t_{ij}' (A,A) [a^\dagger_i(1) a_j(2) + h.c.] +t_{ij}' (A,B) [a^\dagger_i(1) b_j(2) + h.c.] \right. \nonumber \\
&& \left. \;\;\;\;\;
+ t_{ij}' (B,A) [b^\dagger_i(1) a_j(2) + h.c.] +t_{ij}' (B,B) [b^\dagger_i(1) b_j(2) + h.c.]\right] \, .
\end{eqnarray}
Second-order perturbation theory maps ${\cal H}_\perp$ onto a periodic potential
of hexagonal symmetry, similar to the one in Eq.~(3),
%Eq.~(\ref{Hgraphene}) 
with a modulation amplitude $V \simeq 0.06$ eV which we determined numerically
via second order perturbation theory in the interlayer hopping. This energy is smaller than 
$\hbar v_{\rm F} |{\bf G}|/2$ with the Moir\'e superlattice vector ${\bf G}$, regardless
of the rotation  angle between graphene and hBN sheets.

\begin{figure}[h]
\includegraphics[width=12cm]{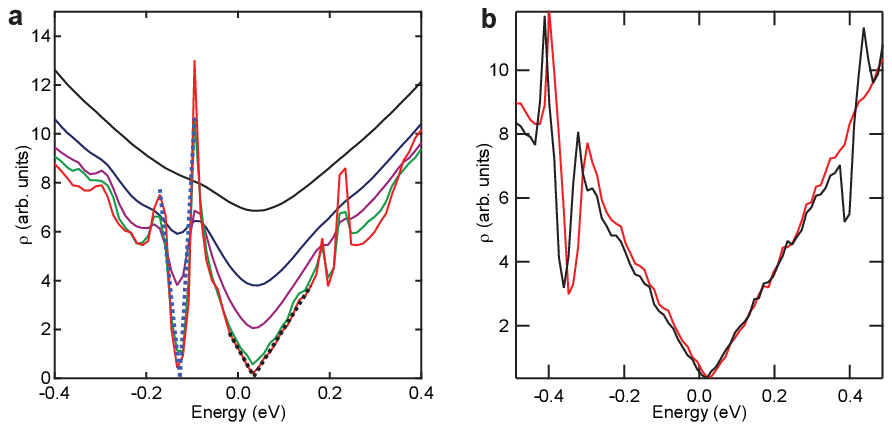} 
\caption{Calculated density of states as a function of coupling. (a) LDOS for a 1000$\times$1000 graphene lattice on a hBN sheet, with relative
rotation $\phi=0.3^o$. The smearing parameter of the Lanczos algorithm is $\zeta=0.09$ (black
curve), 0.04 (red), 0.018 (green), 0.008 (blue) and 0.0036 (violet). The dashed lines give
linear fits close to the Dirac and superlattice Dirac points. (b) LDOS for a 1000$\times$1000 graphene lattice on a hBN sheet, with relative
rotation $\phi=2.0^o$, for $\zeta=0.0036$. The black curve has nearest-neighbor interlayer coupling, the red curve has nearest 
and next nearest
neighbor interlayer hopping.}
\label{fig:tipstrength}
\end{figure}

The total Hamiltonian reads ${\cal H}_1+{\cal H}_2+{\cal H}_\perp$.
We evaluate the LDOS $\rho({\bf r}_i,E)$ on the graphene sheet
using the Lanczos method, which allows to reach linear system sizes of $L=1000$
or more sites~\cite{Dagotto1994a}. The obtained $\rho({\bf r},E)$ depends on a
smearing parameter $\zeta$ which, as long as the density of states of the STM tip and the tunneling rate from the tip to the sample do not depend on energy
can be related to the strength of the tip-graphene coupling.
Figure ~\ref{fig:tipstrength}(a) illustrates how
reducing this coupling allows finer and finer structures in the LDOS to be explored.
For this particular rotation of $\phi=0.3^o$, 
corresponding to a Moir\'e pattern with $\lambda=13.4$ nm, and the set of parameters we just discussed, we see that the LDOS vanishes more or less linearly close to $E-E_{\rm D}
=-\hbar v_{\rm F} |{\bf G}|/2$. Our numerical data suggest a complete, linear vanishing
of the LDOS there.
The linear vanishing of the LDOS, gives an estimate for 
the Fermi velocity close to the three new superlattice Dirac points.
We estimate that the new velocity for the holes is $(0.73\pm0.08)v_{\rm F}$.  The new velocity for %%%%%%%%%%%%%%%
%%%%%%%%%%%%%%%%%%
electrons is smaller because of the increased slope and we estimate it to be $(0.53\pm0.05)v_{\rm F}$.  The new reduced Fermi velocities are in reasonable agreement with our experimentally determined values.

A significant energy asymmetry emerges in that the expected dip
in the density of states is stronger in the valence $E<0$ than in the conduction $E>0$
band. This asymmetry arises because of (i) the asymmetry in the on-site energies in the
hBN layer, which effectively induces a second order potential that is stronger
for negative than for positive energies and (ii) next-nearest neighbor interlayer hopping, which
effectively induces a local periodic modulation of $t_1$. This is illustrated in Fig. ~\ref{fig:tipstrength}(b).

\section{Transport Measurement}
\begin{figure}[h]
\includegraphics[width=8.5cm]{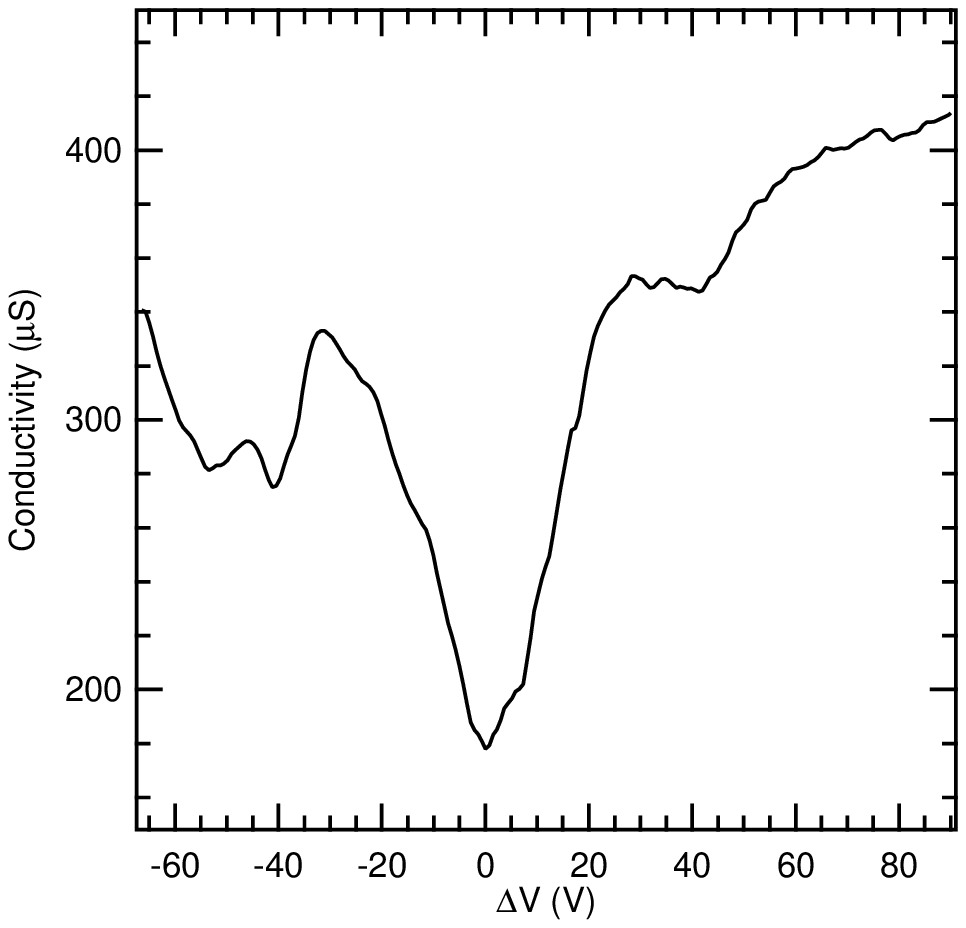} 
\caption{Conductivity as a function of gate voltage for the graphene device with a 13.4 nm Moir\'e pattern.}
\label{fig:transport}
\end{figure}

We have also performed electrical transport measurements on the graphene device shown in Fig. 3.  The conductivity as a function of gate voltage is shown in Fig. ~\ref{fig:transport}.  The gate
voltage is plotted as the offset from the gate voltage at the Dirac point.  We observe two locations of decreased conductivity which are located at approximately $\pm40$ V from the Dirac point.  This is %%%%%%%%%%%%%%%
%%%%%%%%%%%%%%%%%%
the same separation in gate voltage as observed in the spectroscopy measurements, where the main Dirac point was separated from the superlattice Dirac point by 40 V.  Therefore, we conclude that the dips in conductivity are due to the presence of the superlattice Dirac point.  This gives further indirect evidence of the superlattice Dirac point.  We also see a second dip in the conductivity near -50 V.  Our STM topography measurements showed a region of the device with a second Moir\'e pattern of about 10 nm.  The superlattice Dirac point due to this Moir\'e occurs at a higher energy and hence a larger gate voltage.

\end{document}